\documentclass[preprint]{aastex62}
\usepackage{ulem}
\usepackage{natbib}

\received{December 22, 2018}
\revised{...}
\accepted{May 16, 2019}
\submitjournal{ApJ}

\shorttitle{Evolutionary models for UCDs}
\shortauthors{Fernandes et al.}

\begin{document}

\title{\textsc{EVOLUTIONARY MODELS FOR ULTRACOOL DWARFS}}

\correspondingauthor{C. S. Fernandes}
\email{c.fernandes@uliege.be}

\author[0000-0002-0351-2429]{C. S. Fernandes}
\affiliation{Space sciences, Technologies and Astrophysics Research (STAR) Institute, Universit\' e de Li\` ege, 19C All\'ee du 6 Ao\^ ut, B-4000 Li\` ege, Belgium}
\affiliation{Niels Bohr Institute, University of Copenhagen, {\O}ster Voldgade 5, DK-1350 Copenhagen }

\author{V. Van Grootel}
\affiliation{Space sciences, Technologies and Astrophysics Research (STAR) Institute, Universit\' e de Li\` ege, 19C All\'ee du 6 Ao\^ ut, B-4000 Li\` ege, Belgium}

\author{S. J. A. J. Salmon}
\affiliation{Space sciences, Technologies and Astrophysics Research (STAR) Institute, Universit\' e de Li\` ege, 19C All\'ee du 6 Ao\^ ut, B-4000 Li\` ege, Belgium}

\author{B. Aringer}
\affiliation{Dipartimento di Fisica e Astronomia Galileo Galilei, Universit\`a di Padova, Vicolo dell'Osservatorio 3, I-35122 Padova, Italy}
\affiliation{Observatorio Astronomico di Padova - INAF, Vicolo dell'Osservatorio 5, I-35122 Padova, Italy}

\author{A. J. Burgasser}
\affiliation{Center for Astrophysics and Space Science, University of California San Diego, La Jolla, CA, 92093, USA}

\author{R. Scuflaire}
\affiliation{Space sciences, Technologies and Astrophysics Research (STAR) Institute, Universit\' e de Li\` ege, 19C All\'ee du 6 Ao\^ ut, B-4000 Li\` ege, Belgium}

\author{P. Brassard}
\affiliation{D\'epartement de Physique, Universit\'e de Montr\'eal, Montr\'eal, Qu\'ebec H3C 3J7, Canada}
\author{G. Fontaine}
\affiliation{D\'epartement de Physique, Universit\'e de Montr\'eal, Montr\'eal, Qu\'ebec H3C 3J7, Canada}



\begin{abstract}

Ultracool dwarfs have emerged as key targets for searches of transiting exoplanets. Precise estimates of the host parameters (including mass, age, and radius) are fundamental to constrain the physical properties of orbiting exoplanets. We have extended our evolutionary code CLES (Code Li\'egeois d'Evolution Stellaire) to the ultracool dwarf regime. We include relevant equations of state for H, He, as well as C and O elements to cover the temperature-density regime of ultracool dwarf interiors. For various metallicities, we couple the interior models to two sets of model atmospheres as surface boundary conditions. We show that including C and O in the EOS has a significant effect close the H-burning limit mass. The typical systematic error associated with uncertainties in input physics in evolutionary models is $\sim 0.0005 M_\odot$. We test model results against observations for objects whose parameters have been determined from independent techniques. We are able to reproduce dynamical mass measurements of LSPM J1314+1320AB within $1\sigma$ with the condition of varying the metallicity (determined from calibrations) up to $2.5\sigma$. For GJ 65AB, a $2\sigma$ agreement is obtained between  individual masses from differential astrometry and those from evolutionary models. We provide tables of ultracool dwarf models for various masses and metallicities that can be used as reference when estimating parameters for ultracool objects.

\end{abstract}



\section{Introduction}
\label{sect1}
Ultracool dwarfs (UCDs) are cool, small and dim objects lying at the faint, red end of the Hertzsprung-Russell (HR) diagram (M7 and cooler; \citealt{Kirkpatrick1995}), at the limit of core H-burning. UCD classifications encompasses the very low-masses stars (VLMS) and brown dwarfs (BDs). 
As these objects emit primarly in the near-infrared (NIR) and have low intrinsic luminosities, 
it took the development of efficient photometric detectors and spectrographs at NIR wavelengths in the late 1990s to populate the UCD region in the HR diagram (see sect. 2.2 in \citealt{Reid2005} for a historical review). Today, it is known that UCDs represent at least 15\% of the population of star-like objects in the solar neighborhood \citep[see, e.g.,][and the RECONS collaboration\footnote{See \url{www.recons.org}}]{Bartlett2017}. 

In parallel with the increased observational access to UCDs, theoretical evolutionary models have improved in the past decades. Key advances include \citet{DAntona1996} who generated models for metal-poor stars, \citet[][the "Tucson group"]{Burrows1993,Burrows1997} who included grain opacities in atmosphere models, and \citet[][the "Lyon group"]{Baraffe1995,Baraffe1998} who coupled interior models to full model atmospheres as boundary conditions. The most recent and commonly used model set is \cite{Baraffe2015} (hereafter, BHAC15), who presented updated evolutionary models for low-mass stars and young BDs that use boundary conditions from BT-Settl model atmospheres \citep{Allard2012a,Allard2012b, Rajpurohit2013, Rajpurohit2018}. These atmospheres include updated molecular opacity linelists and cloud formation, as well as atmospheric convection parameters calibrated with 2D/3D radiative hydrodynamic simulations \citep{Freytag2010,Freytag2012}. BHAC15 models also adopt the solar composition of \citet{Asplund2009}, supplemented with abundances from \citet{Caffau2011} for C, N, O, Ne, P, S, K, Fe, Eu, Hf,Os, and Th. 
The BHAC15 models provide a significant improvement over the \citet{Baraffe1998} models when comparing to observations. In particular, the weaknesses in the \citet{Baraffe1998} colour-magnitude diagrams (e.g., for the optical ($V-I$) colors that were too blue for a given magnitude; and for the NIR colors for relatively old objects) were significantly improved by BHAC15. 

The BHAC15 grid is publicly available and is widely used when trying to retrieve the parameters of UCDs (e.g. age, mass) from observations. This grid is given for solar composition with stellar mass from $1.40 M_\odot$ down to $0.01 M_\odot$, with the lowest $T_{\rm eff} = 1200$ K for 0.01 $M_\odot$, which corresponds to an age of 41 Myr. At low effective temperatures, the limit of validity of evolutionary models is set by the limit of validity of model atmospheres \citep{Saumon2008}.
These models are likely robust for $T_{\rm eff} \gtrsim 2000$ K, as below this temperature there are significant uncertainties associated with cloud formation and condensation of chemical compounds which become important opacity sources in the atmosphere \citep{Tsuji1996, Marley2009, Morley2012}.

In this paper, we present adaptations made to our in-house evolutionary code CLES (Code Li\'egeois d'Evolution Stellaire; Scuflaire et al. 2008) to produce UCD models: relevant equations-of-state (EOS) suitable for the low temperature and high density regime, and relevant model atmospheres used as boundary conditions for the interior. The main motivation behind this work is the SPECULOOS project (Search for Planets EClipsing ULtra-cOOl Stars; \citealt{Gillon2017b, delrez2018speculoos}), a survey searching for transiting planets in the habitable zone of the nearest and brightest UCDs. The prototype of the SPECULOOS project is the TRAPPIST telescope (TRAnsiting Planet and PlanetesImals Small Telescope; \citealt{Gillon2011}) which led to the discovery of seven Earth-sized planets transiting the ultracool dwarf TRAPPIST-1 \citep{Gillon2016,Gillon2017a}. More generally, transiting surveys are now focused on very cool stars, because of their small radii with close-in habitable zones. Atmospheric characterization of these exoplanets will be within reach of the next-generation telescopes (e.g. James Webb Space Telescope, European Extreme Large Telescope). In this context, precise estimates of the mass, radius, luminosity, effective temperature and age for a host star are important to thoroughly characterize its exoplanets \citep[see, for TRAPPIST-1,][]{VanGrootel2018}.

This paper is organized as follows. Section \ref{sect2} provides the details of our CLES evolutionary models for UCDs, general properties and systematic error estimate, and comparison to BHAC15 UCD models. Section \ref{sect3} gives a series of test-cases for model comparison to observations. Section \ref{sect4} presents our conclusions. 
\section{CLES models for ultracool dwarfs}
\label{sect2}
\subsection{Input physics}
\label{std}
The CLES evolutionary code was developed in the early 2000s by the asteroseismology group of the University of Li\`ege, and has been continuously updated. We present here only the main input physics used for UCDs, and refer to \cite{Scuflaire2008} for the main numerical features \citep[see also][for CLES in the context of the Sun and solar-like stars]{Buldgen2016,Buldgen2017a,Buldgen2017b}. CLES includes different choices for EOS, opacity and atmosphere tables but here we mention only those relevant for the study of UCDs. 

We considered H, He, C, and O for the EOS of UCD objects. We directly adapted tables built for white dwarfs and subdwarf B stars \citep{Brassard1994}. These tables cover a large domain of the temperature-density plane ($2.10 \leq \log T \leq 8.98$ and $-12.0 \leq \log \rho \leq 9.0$), which includes UCDs at all evolutionary stages. Four EOS tables are available, one for each element considered. For each table, three regimes are invoked. First, for the low-density region, a network of Saha equations is solved for a mixture of radiation and an almost ideal (including Coulomb corrections), non-degenerate, partially or fully ionized gas composed of a mixture of H, He, C, and O in various proportions. Second, in the partial ionization region where non-ideal and electron degeneracy effects are important (intermediate densities), we used the EOS of \citet{Saumon1995} for H and He, an improved version of the EOS of \citet{Fontaine1977} for C and for O. 
Third, the high-density domain corresponds to the fully ionized plasma in liquid and, ultimately, solid phases according to the physics described in \citet{Lamb1974} and improved by \citet{Kitsikis2005}.
The low-density boundary of the second regime matches very smoothly with the high-density boundary of the first regime, thus ensuring that there are no significant jumps in the thermodynamic variables of interest. The connection between the high-density boundary of the partial degeneracy regime with the low-density boundary of the totally ionized domain is made at the location where the electron degeneracy parameter $\eta$ is equal to 20. Care has been taken to insure that thermodynamic consistency is respected as explained in \citet{Fontaine1977}. 
Interpolation in composition is handled following the additive volume prescription of \citet{Fontaine1977}.
This additive volume prescription is the best viable option for handling mixture in the EOS of individual elements \citep{Vorberger2007,Wang2013,Danel2015}. The FreeEOS equation of state\footnote{See \url{http://freeeos.sourceforge.net/} by A. Irwin.} \citep{Irwin2012} version 2.2.1. (EOS1 configuration) is also available in CLES. It calculates the EOS without radiation pressure for specified mixtures using an efficient free-energy minimization technique. However, this did not converge for the lowest temperatures/highest densities encountered in H-burning stars below $\sim0.14$ $M_{\odot}$ at solar composition.

Model atmospheres based on $T(\tau)$ relations (including grey atmospheres) are inaccurate as boundary conditions (BCs) for interior models of UCDs \citep{Chabrier1997}. 
We have implemented in CLES two sets of surface boundary conditions (density, temperature, geometrical depth) from detailed model atmospheres. The first derives from the publicly available BT-Settl model atmospheres \citep{Allard2012a,Allard2012b}. The second makes use of the model atmospheres of B. Aringer, originally developed for Asymptotic and Red Giant Branch stars \citep[][hereafter, AR16]{Aringer2016}. In both cases, the transition between interior and atmosphere is performed at the Rosseland mean optical depth $\tau = 100$ (similar to the BHAC15 models), as a safe limit to avoid discrepancies at the boundary while treating convection and adiabatic processes \citep{Chabrier1997}. 
For BT-Settl model atmospheres and interior structure, we used the models computed with the solar abundances of \cite{Asplund2009}, as for the interior structure, and followed the same heavy-element abundance as for the Sun (i.e., $[\alpha/H] =0$). This results in four distinct solar-scaled compositions: $[M/H]=-0.5, 0.0, +0.3, +0.5$.
The AR16 models are computed for \cite{Asplund2009} solar (and meteoritic when available) abundances as well, supplemented by \cite{Caffau2011} for C, N, O, Ne and Ar. Seven compositions are possible, from $[M/H]=-2$ to 1, by steps of 0.5. AR16 is computed with the COMARCS program which is based on the version of the MARCS code of \cite{Gustafsson2008}, and uses the COMA (Copenhagen Opacities for Model Atmospheres) opacity generation code  by \cite{Aringer2000}. Unlike the BT-Settl models, AR16 models are dust-free and thus limited to $T_{\rm eff} \geq$ 2600 K.

CLES includes opacities from the OPAL \citep{Iglesias1996} or the OP \citep{Badnell2005} project.
In both cases, these opacities are combined for low temperatures to opacities from \cite{Ferguson2005}. The effects of thermal conductivity have been taken into account following \cite{Potekhin1999} and \cite{Cassisi2007}. 

Nuclear reaction rates for the pp chain come from the review of \citet{Adelberger2011}, except for the $^7$Li$(p,\alpha)^4$He reaction which comes from the NACRE II compilation \citep{Xu2013}. 

Convection is treated using the mixing length theory (MLT). For UCDs, we generally set $\alpha_{MLT}$ (the ratio between the mixing length and the pressure scale height) to $\sim2.0$, the value adopted in BT-Settl model atmospheres for such stars (BHAC15). AR16 model atmospheres also follow the MLT formalism, with $\alpha_{MLT} = 1.5$, which we adopt for the interior when using these models for surface boundary condition. In Sect. \ref{syst} we examine the effect of $\alpha_{MLT}$ on the evolution of UCDs.
 
CLES solar calibration (evolutionary track giving the Sun luminosity and effective temperature at its present age), without diffusion, with OPAL opacities, with our standard H+He+C+O EOS, and BCs from BT-Settl model atmospheres, gives $\alpha_{MLT} = 1.8$, $X_{0} = 0.729$, and $Z_{0} = 0.013$. This is our standard CLES configuration for UCDs (with $\alpha_{MLT} = 2.0$). We note that this solar calibration does not depend highly on the chosen input physics, that is model atmospheres, EOS, solar abundances \citealt{Asplund2009}/\citealt{Caffau2011}, and opacities.

\subsection{Properties of CLES models}
\subsubsection{H-burning limit mass}
\label{Mhbl}
\begin{figure}[!ht]
\begin{center}
\includegraphics[scale=0.4,angle=0]{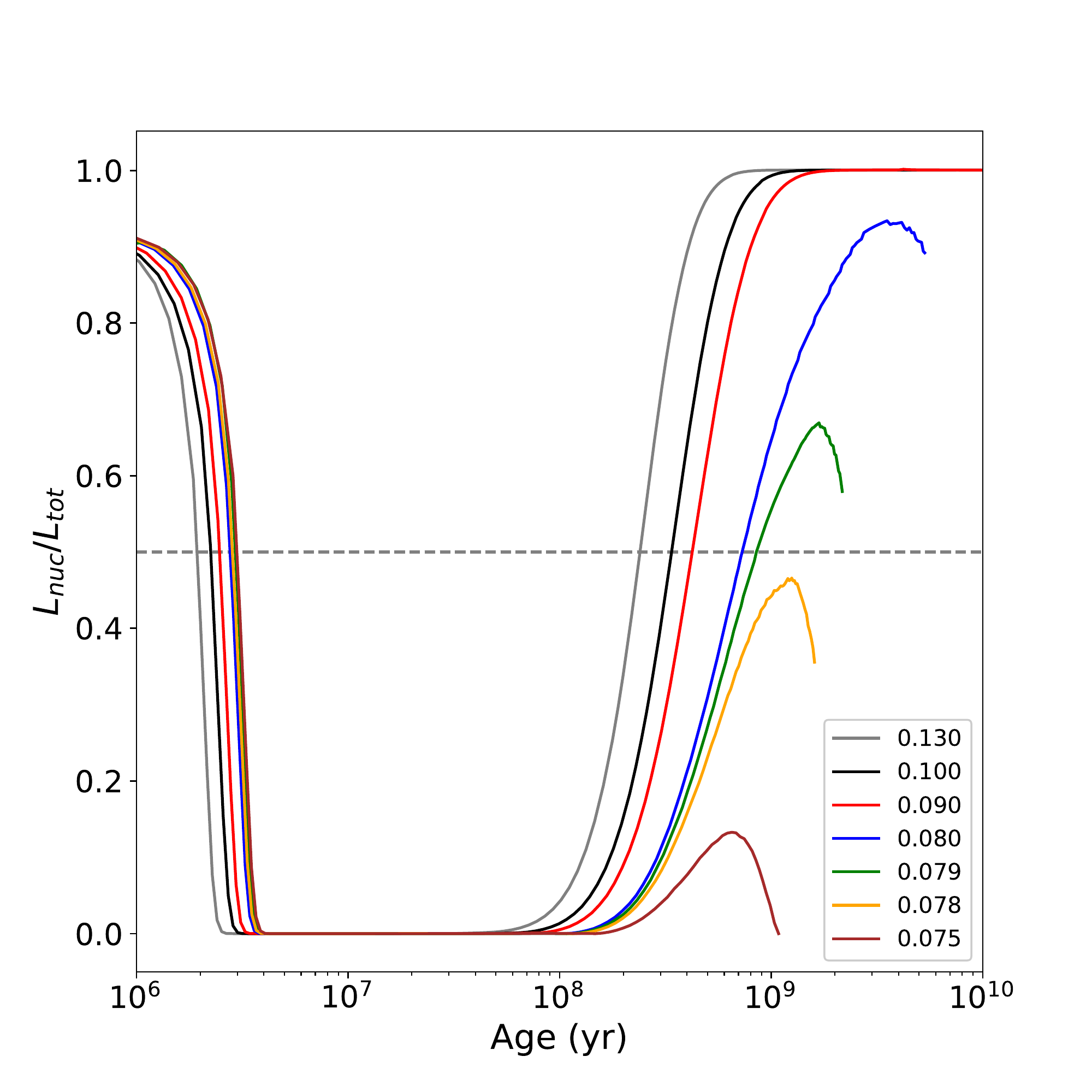}
\end{center}
\caption{\label{fig:2-Lnuc} The evolution of fractional luminosity contributed by nuclear fusion reactions $L_{nuc} / L_{total}$ for UCD CLES models, showing the tracks between 0.130 and 0.075 $M_{\odot}$ assuming solar composition. The dashed line marks the 50\% limit set in \cite{Reid2005} for formally defining $M_{HBL}$, which in CLES is slightly above 0.078 $M_\odot$.} 
\end{figure}

The formal H-burning limit mass ($M_{HBL}$), i.e., the mass where the fractional contribution to the total luminosity due to hydrogen fusion  $L_{nuc} / L_{total}$ never exceeds 50\% \citep{Reid2005}, is slightly above 0.078 $M_{\odot}$  in standard CLES configuration (Fig. \ref{fig:2-Lnuc}).  $M_{HBL}$ is about 0.07 $M_{\odot}$ in the BHAC15 models according to their public grids, and 0.073 $M_{\odot}$ for \citet{Burrows1993,Burrows1997} models. 
Figure \ref{fig:2-Lnuc} shows that stars with mass slightly higher than 0.08 $M_{\odot}$ achieve stable luminosities and temperatures for many Hubble times, while $0.080 M_\odot$ and $0.079 M_\odot$ objects are transition objects, able to sustain fusion for several hundreds of million years, but eventually cooling degenerately as brown dwarfs. The region where $L_{nuc} / L_{total} > 0.5$ at early ages for all masses in Fig. \ref{fig:2-Lnuc} corresponds to the short-lived phase of D-burning. The $M_{HBL}$ depends on the chemical composition of the star such that the mass limit increases as metallicity decreases \citep[see also section 3.4 in][]{Reid2005}. For example, in the CLES models the $M_{HBL}$ is 0.074, 0.078 and 0.080  $M_{\odot}$ at $[M/H]=+0.5, +0.0$ and $-0.3$, respectively.

\subsubsection{Abundance of light elements}
The fusion of light elements, namely Li, Be, B, can occur in the interior of UCDs depending on their mass and age. Presence/absence of light elements in the spectra provide a powerful diagnostic to distinguish UCDs of various ages and masses, including BDs (\citealt{Chabrier1997}). For example, observing the lithium doublet at $\lambda LiI=6708 \AA$ was proposed by \citet{Rebolo1992} as the famous \textit{lithium test} that confirmed the first BD candidates \citep{Rebolo1995, Basri1996}.

\begin{figure}[!ht]
\begin{center}
\includegraphics[scale=0.4,angle=0]{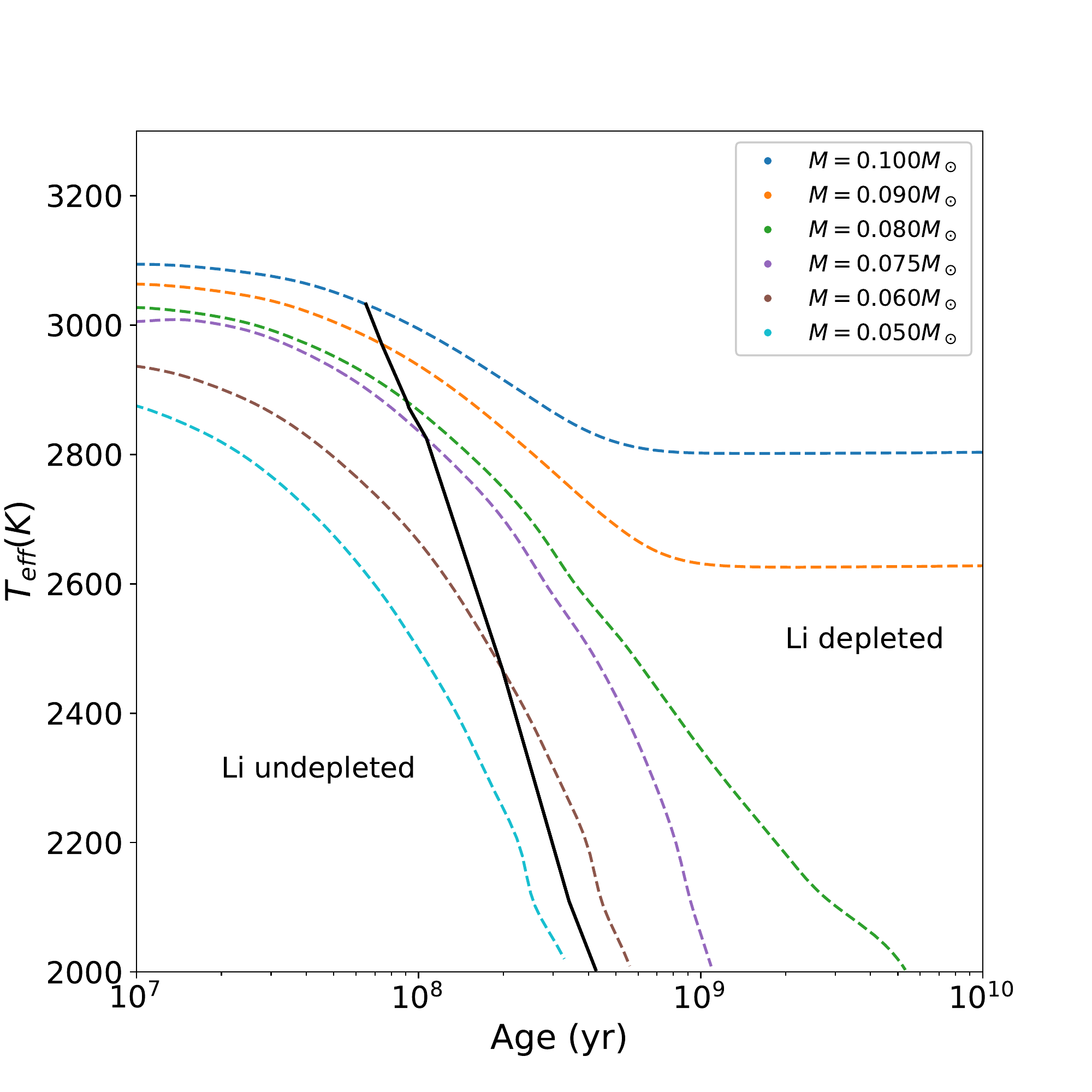}
\end{center}
\caption{\label{fig:2-Lithium} The evolution of effective temperature in CLES models, showing the tracks from 0.100 to 0.050 $M_{\odot}$ assuming solar composition. Models to the right of the solid line have depleted lithium by over $90\%$ from its initial abundance.} 
\end{figure}

We computed the minimum burning mass for $^7$Li, $^9$Be and $^{10}$B in CLES models to be 0.053 $M_\odot$, 0.065 $M_\odot$ and 0.079 $M_\odot$, respectively, in agreement with literature values, e.g., 0.055 $M_\odot$, 0.065 $M_\odot$ and 0.08 $M_\odot$, respectively, for \cite{Chabrier1997}, and of about 0.055 $M_\odot$, 0.07 $M_\odot$ and 0.09 $M_\odot$, respectively, for \cite{Burrows1997}. Fig. \ref{fig:2-Lithium} shows the boundary for lithium depletion for when $90\%$ of lithium has been consumed for different masses.

\subsection{Systematic error in evolutionary models}
\label{syst}
\subsubsection{The choice of EOS}

\begin{figure}[!ht]
\begin{center}
\includegraphics[scale=0.4,angle=0]{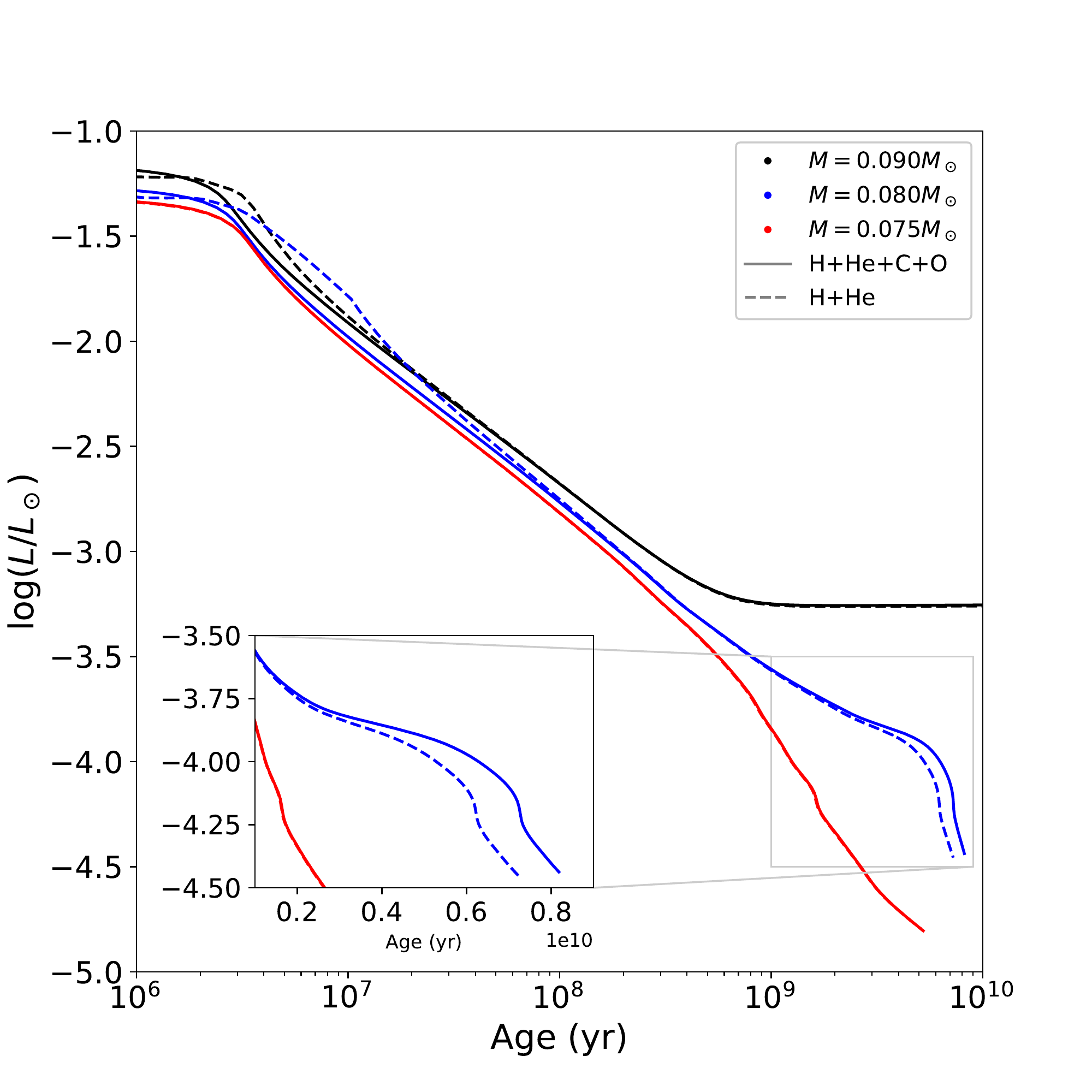}
\includegraphics[scale=0.4,angle=0]{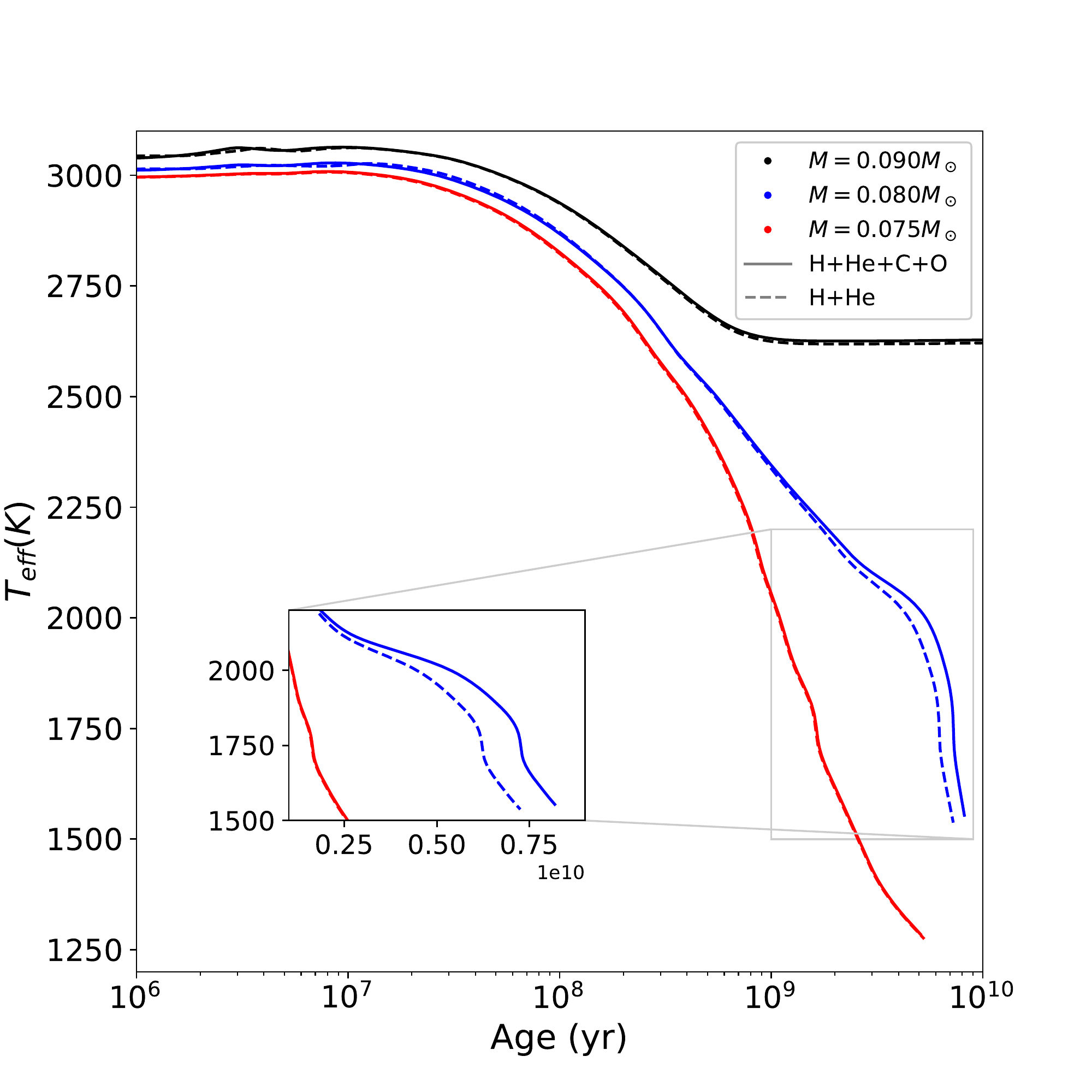}
\caption{
\label{fig:2-EOSa} Stellar luminosity (left) and effective temperature (right) as a function of age for 0.075 $M_{\odot}$ (below $M_{HBL}$), 0.08 $M_{\odot}$ (at $M_{HBL}$), and 0.09 $M_{\odot}$ stars at solar composition, comparing CLES standard EOS (solid line) and H+He EOS where metals have been assimilated to Helium (dashed line). Close to (far from) $M_{HBL}$, differences in luminosity and effective temperature are about 4\% (1\%) and 1\% (0.2\%), respectively. 
}
\end{center}
\end{figure}

\begin{figure}[!ht]
\begin{center}
\includegraphics[scale=0.4,angle=0]{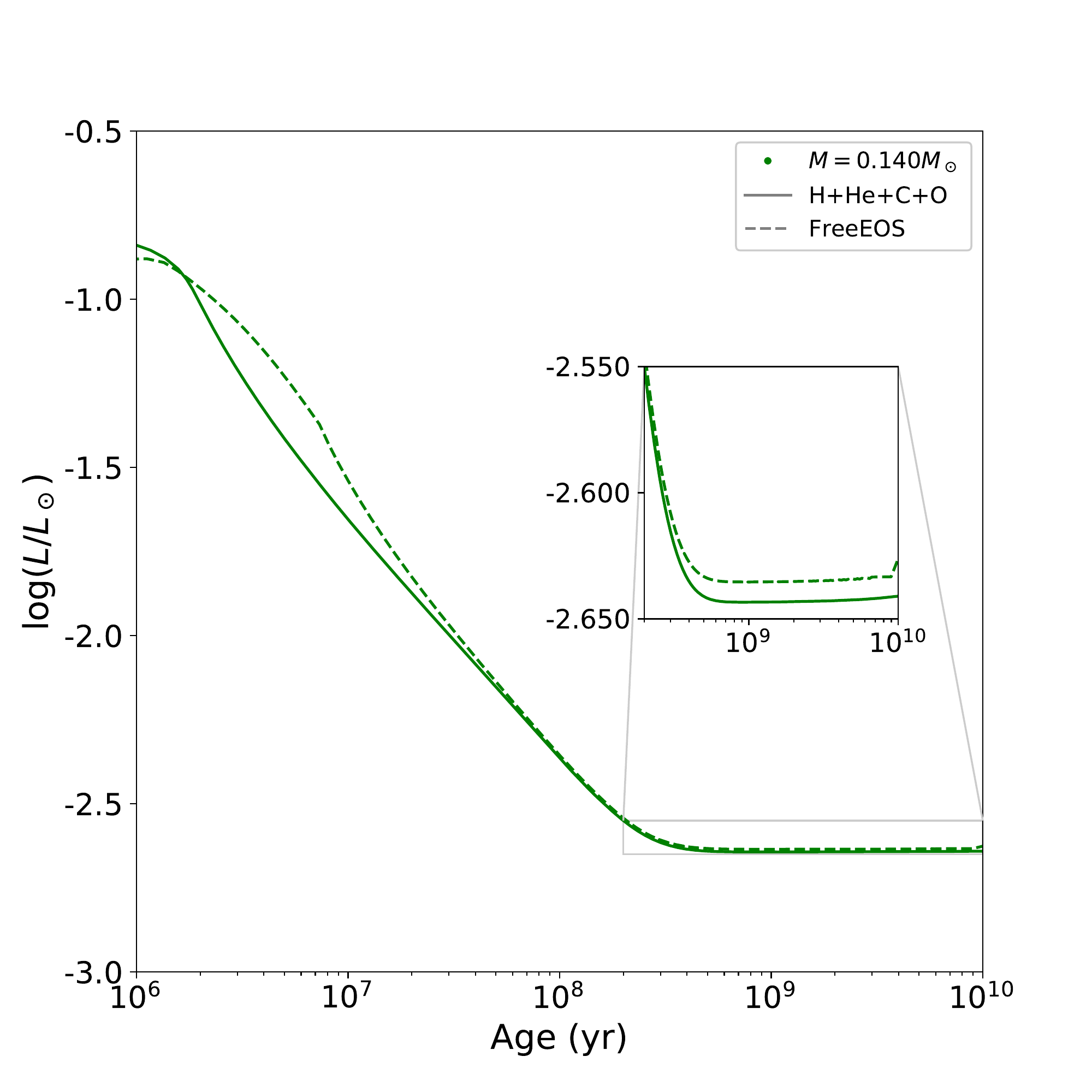}
\includegraphics[scale=0.4,angle=0]{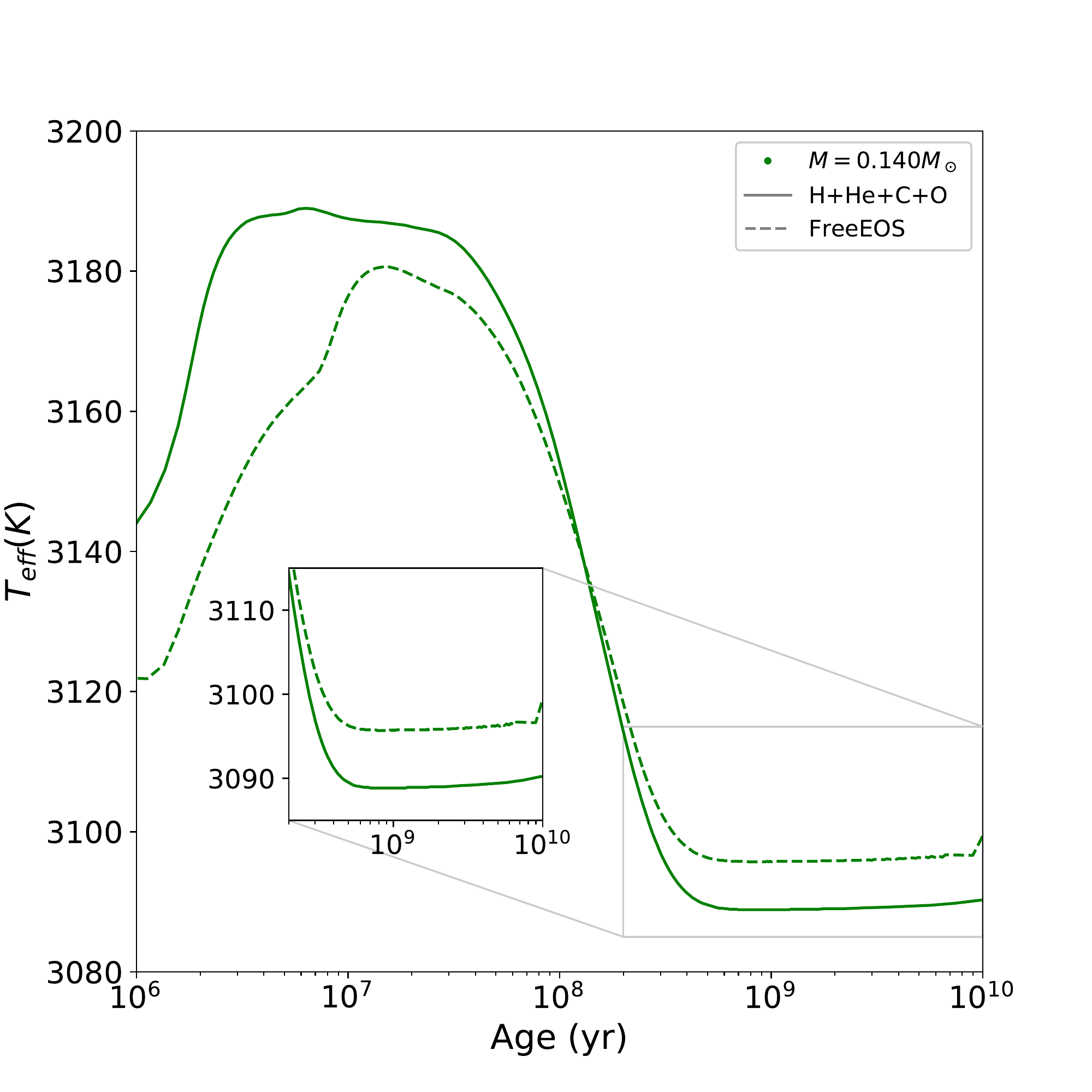}
\end{center}
\caption{\label{fig:2bis-EOSb} Stellar luminosity (left) and effective temperature (right) as a function of age for 0.14 $M_{\odot}$ stars at solar composition, comparing CLES standard EOS (solid line) and FreeEOS EOS (dashed line): difference at canonical age of about 2\% in luminosity and 0.2\% in effective temperature. 
}  
\end{figure}

In our standard EOS, C and O are proxies for all metals, with the same C/O proportion as in \cite{Asplund2009}. BHAC15 (and older generation) models gather all metals, including C and O, in an increased fraction of He.\footnote{The mixture of H EOS and He EOS is also handled by the additive volume prescription in BHAC15 and older generation models.} This is only valid when ionic pressure is negligible compared to electronic pressure, which is true through most of a UCD interior \citep{Chabrier1997}. Figure \ref{fig:2-EOSa} quantifies this effect for the first time. We compare stellar luminosity and effective temperature as a function of age, at solar composition, for 0.075 $M_{\odot}$ (below $M_{HBL}$), 0.08 $M_{\odot}$ (close to $M_{HBL}$), and 0.09 $M_{\odot}$ (above $M_{HBL}$) stars with the CLES standard configuration EOS (H+He+C+O; solid lines) and the EOS used in the BHAC15 models (H+He; dashed lines). Figure \ref{fig:2-EOSa} shows that including the EOS for C and O has a strong effect at and close to $M_{HBL}$, with diverging evolutionary tracks at 0.08$M_{\odot}$. Close to the $M_{HBL}$ (0.079 and 0.081 $M_\odot$), the effect on the luminosity and effective temperature is $+4\%$ and $+1\%$, respectively. 
Whereas farther from $M_{HBL}$ (0.09$M_{\odot}$ and higher masses; 0.075$M_{\odot}$ and lower), the impact is smaller, about $+1\%$ in luminosity and $+0.3\%$ in effective temperature. The conclusion of this experiment is that assimilating all of the metals into He is generally a valid hypothesis, but care must be taken close to the $M_{HBL}$.
Some modern stellar evolution codes (e.g. DARTMOUTH, \citealt{Feiden2016}, and references therein; and PARSEC, \citealt{Chen2014}), used to model very low-mass stars, provide public grids down to $\sim$ 0.10 $M_{\odot}$, based on the FreeEOS EOS. 
We found it impossible to make FreeEOS converge for densities/temperatures corresponding to stellar masses below $\sim$ 0.14 $M_{\odot}$.  It is plausible that the differences in minimum mass found by different groups come from choosing different FreeEOS configurations when computing its tables, in particular regarding the numerous ionization states of various elements.
In Fig. \ref{fig:2bis-EOSb}, we show the evolution in luminosity of a 0.14 $M_{\odot}$ star with FreeEOS and our standard EOS. Differences observed at canonical age are about $2\%$ in luminosity and $0.2\%$ in effective temperature.

\subsubsection{The choice of model atmospheres as BCs}
\begin{figure}[!ht]
\begin{center}
\includegraphics[scale=0.4,angle=0]{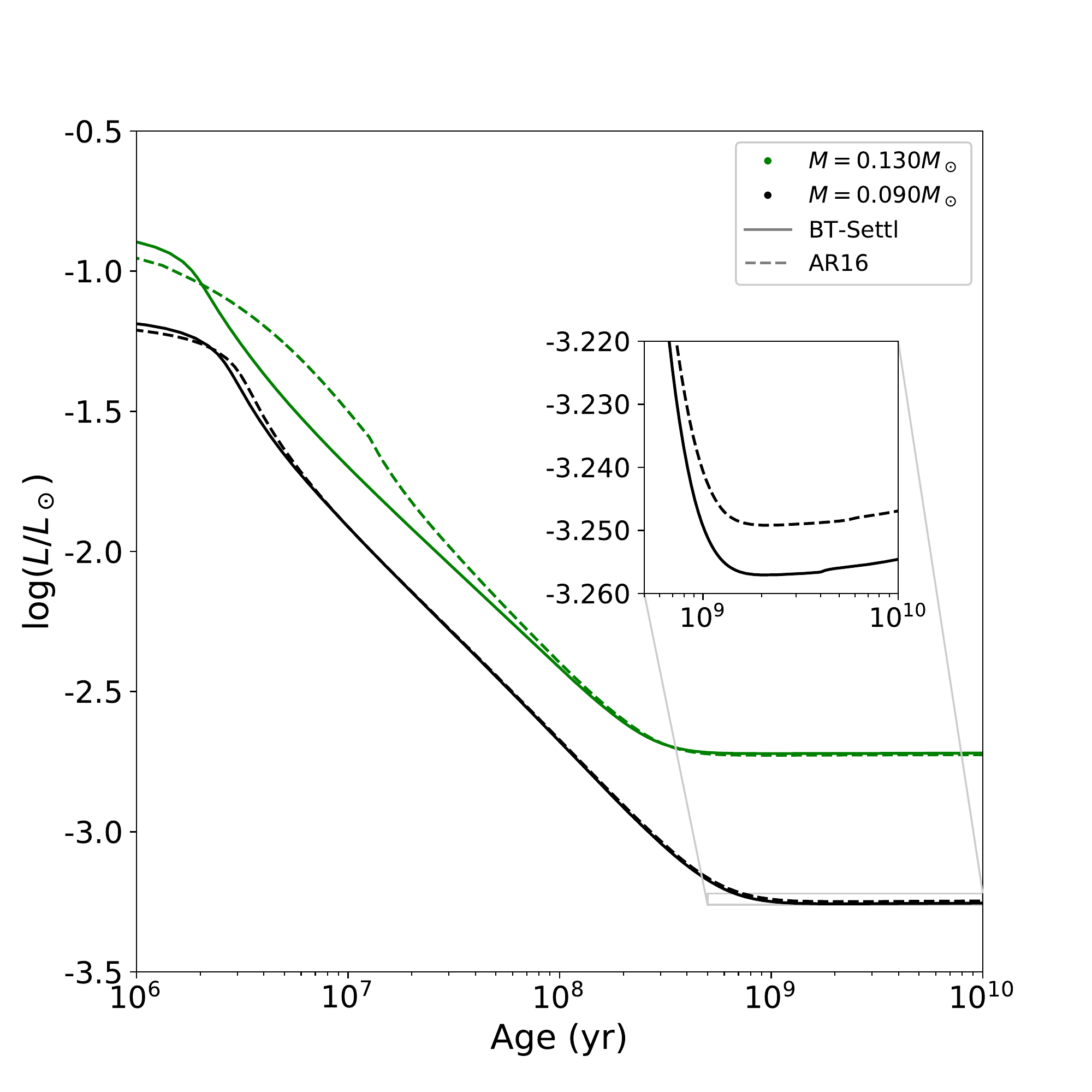}
\includegraphics[scale=0.4,angle=0]{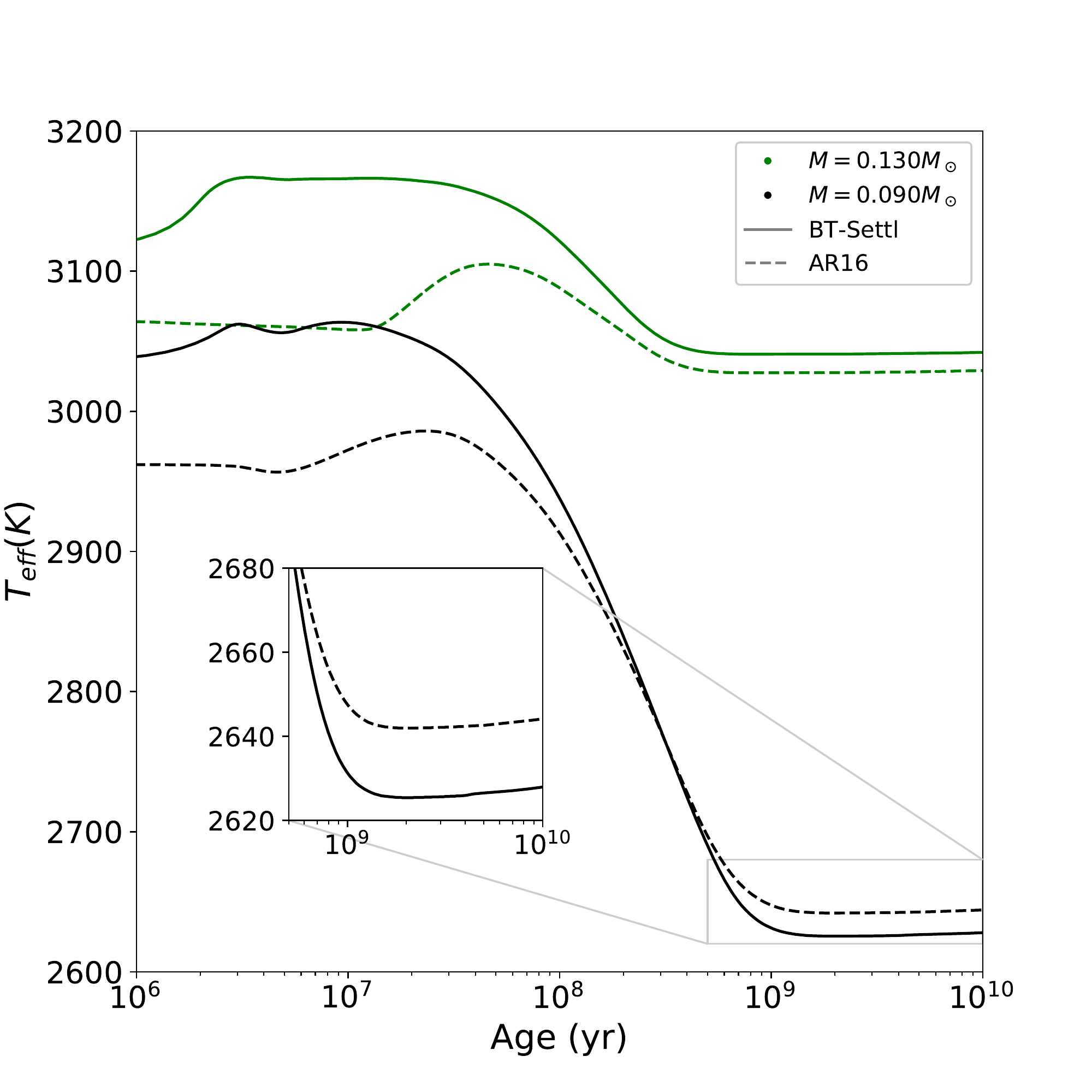}
\end{center}
\caption{
\label{fig:2-BAR}
Stellar luminosity (left) and effective temperature (right) as function of age for 0.09 $M_{\odot}$ and 0.13 $M_{\odot}$ stars at solar composition, comparing BCs extracted from BT-Settl model atmospheres (solid line), and those of AR16 (dashed line). Maximum difference in models with and without grain formation (BT-Settl and AR16, respectively): about $1-2\%$ in luminosity and $0.4-0.7\%$ in effective temperature. 
}  
\end{figure}

Figure \ref{fig:2-BAR} compares luminosity and effective temperature as a function of age, at solar composition, for 0.09 $M_{\odot}$ and 0.13 $M_{\odot}$ with BCs extracted from BT-Settl model atmospheres and AR16 models. Systematic error is small in this case, with maximum differences (in H-burning phase) of $1-2\%$ in luminosity, $0.4-0.7\%$ in effective temperature, and $0.2-0.4\%$ in radius. Models built with BCs based on BT-Settl atmospheres, which include grain formation below $T_{\rm eff} <$ 2600 K, are used in a more extended range of stellar mass (including young BDs) than those based on AR16 models, but the latter is more extended in terms of metallicities. 

\subsubsection{Other sources of systematic error}
We computed evolutionary tracks with OPAL and OP opacity tables for various masses and various ages at solar composition. OP and OPAL tables give very similar results, i.e., no systematic error is observed by choosing OP rather than OPAL.

We also computed evolutionary tracks with \citet{Caughlan1988} nuclear reaction rates, which were implemented in some older versions of CLES. We found a typical $+1\%$ increase in luminosity for 0.09 $M_{\odot}$ at 10 Gyr using \citet{Caughlan1988} instead of \citet{Adelberger2011}. 

It is difficult to give estimates linked to the 'imperfect' description of convection by the mixing length theory. We note that for UCD objects, the exact value of $\alpha_{MLT}$ (1.5, 2.0 or 2.2) has very little influence on model global parameters such as luminosity, radius, and effective temperature, implying that systematic error linked to MLT theory is negligible. 
\cite{Chabrier2000} also reports that varying $\alpha_{MLT}$ value between 1 and 2 in the interior has no impact for stars below 0.60 $M_\odot$ (see also \citeauthor{Montalban2000} \citeyear{Montalban2000}). We observed the same in our models of UCDs. Only very small values of $\alpha_{MLT}$ significantly changes UCD parameters (see section \ref{sect3.2}).

\subsubsection{Systematic error in models: conclusions}
In order to provide estimates of the typical systematic error associated to evolutionary models, we carried out a series of experiments reproducing the traditional use of evolutionary models by observers, which consists in inferring a mass from luminosity measurements (determined from parallax and spectral energy distributions typically; see \citealt{Filippazzo2015}). In order to 'translate' systematic error in luminosity by changing input physics (EOS, model atmospheres, nuclear reaction rates) into a systematic error in mass, we identified which shift in mass (for models of a given age) would give a shift in luminosity that is typical for the uncertainty in the measurement of luminosity. We found that all systematic shifts observed in luminosity correspond to a modest change in mass, typically between 0.0001 and 0.0005 $M_{\odot}$. We propose that the typical systematic error in mass from evolutionary models, in the range of mass of UCD objects, is 0.0005 $M_{\odot}$. 

\subsection{Comparison to existing UCD models}
\label{compareBHAC}
\begin{figure}[!ht]
\begin{center}
\begin{tabular}{lll}
\includegraphics[scale=0.3,angle=0]{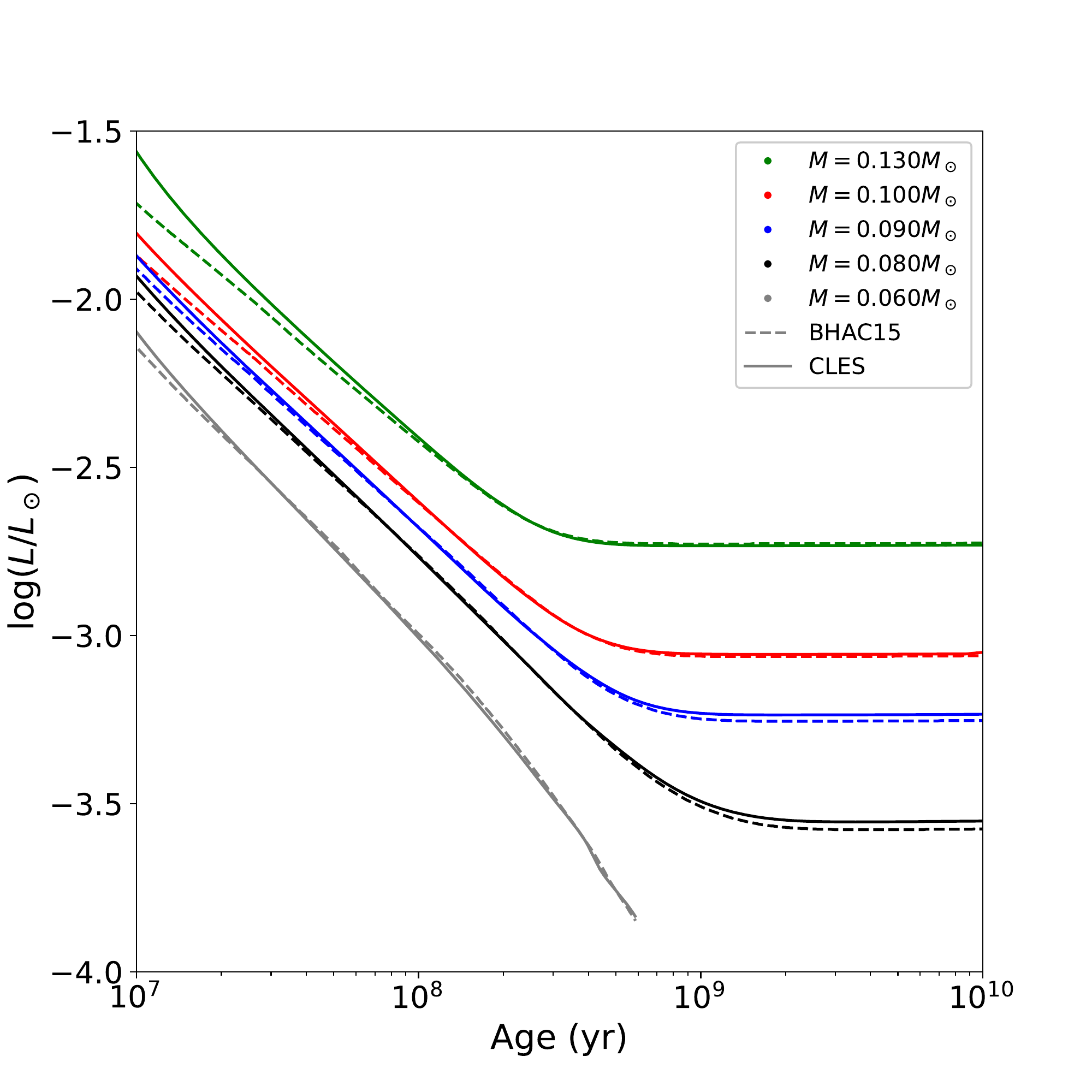} &
\includegraphics[scale=0.3,angle=0]{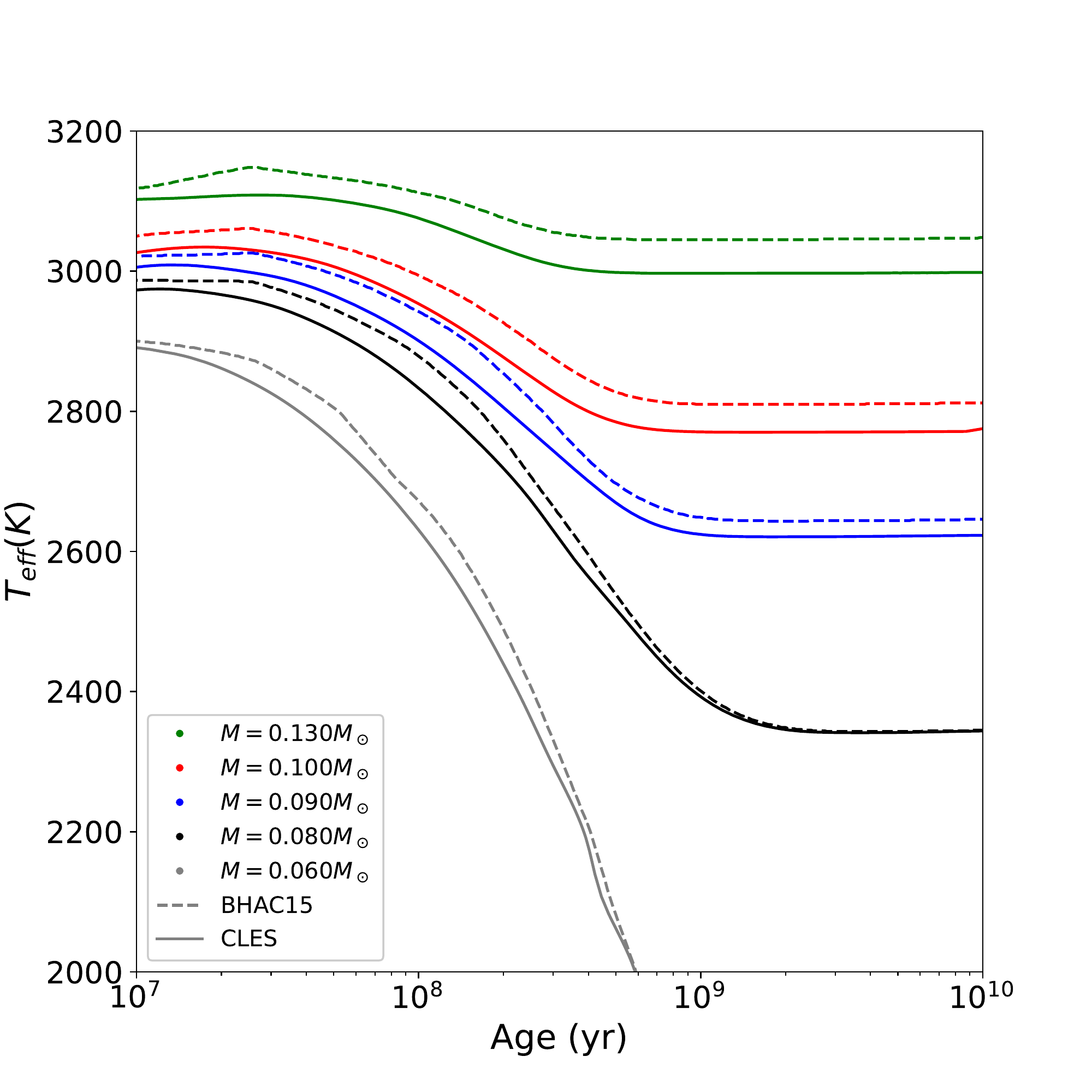} &
\includegraphics[scale=0.3,angle=0]{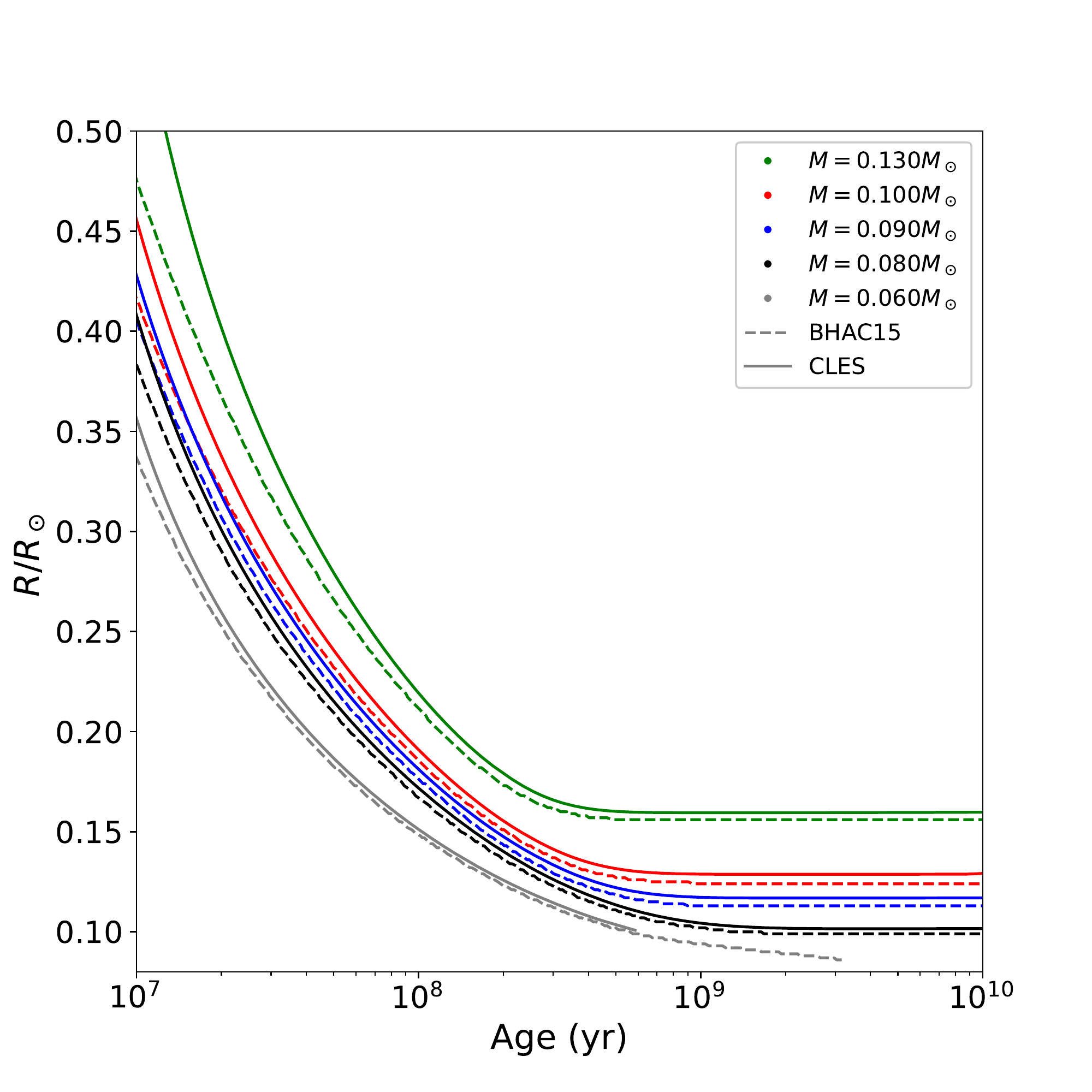}
\end{tabular}
\end{center}
\caption{\label{fig:2-CLES_BHAC15}Comparison between CLES (solid lines) and BHAC15 (dashed lines) models for similar input physics, 0.06, 0.08, 0.09, 0.10, and 0.13 $M_{\odot}$ (grey, black, blue, red, and green, respectively) showing luminosity (left), effective temperature (middle), and radius (right). For the same age, maximum difference between models with similar input physics is of the order of 1-3\%.}
\end{figure}

The BHAC15 models are commonly used when characterizing ultracool dwarfs. To compare CLES with BHAC15 models, we have adapted, as far as possible, identical input physics at solar metallicity: \citet{Grevesse1993} abundances for the interior models and \citet{Asplund2009} supplemented by \citet{Caffau2011} for some elements for BCs from BT-Settl model atmospheres; EOS for H and He only, with an increased fraction of He for assimilating metals; identical initial composition (I. Baraffe, priv. comm.); OPAL opacities, no diffusion, $\alpha_{MLT}=1.6$. 
A comparison between CLES models and those of BHAC15 with identical input physics has already been presented in \cite{VanGrootel2018} with evolutionary tracks for 0.08, 0.09 and 0.10 $M_\odot$ stars. Both models have similar luminosity and differences in effective temperature and stellar radius are of about 1\% and 3\%, respectively. We complement this study by also including tracks for 0.130 and 0.06 $M_\odot$ (the latter below $M_{HBL}$), seen here in figure \ref{fig:2-CLES_BHAC15}. 

\section{Test cases}
\label{sect3}
We test our CLES models by comparing theoretical results with observations for objects for which stellar parameters (luminosity, dynamical mass, age and/or inferred radius from interferometry) have been estimated from independent techniques: the M7 binary LSPM J1314+1320AB \citep[][hereafter D16]{Dupuy2016}, and the spectroscopic twins GJ 65AB and Proxima Centauri \citep[][hereafter K16]{Kervella2016}. We also compare these results with those from BHAC15 models.

\subsection{LSPM J1314+1320AB}
\label{sect3.1}
LSPM J1314+1320AB is a pre-main sequence, nearby binary which has precise measurements of the total dynamical mass and integrated luminosity ($M_{tot}=0.1761 \pm 0.0015 M_\odot$ and $\log(L_{tot,bol}/L_\odot) =  -2.322 \pm 0.009$, respectively), as well as the individual masses and luminosities within $\lesssim1\%$ precision, obtained through spatially-resolved absolute and relative astrometric monitoring and optical and NIR photometry and spectroscopy (D16). The metallicity of LSPM J1314+1320AB has been estimated by D16, using the calibration of \citet{2014AJ....147..160M}, to [Fe/H]=0.04$\pm$0.08, thus consistent with solar metallicity.

In a first test, we used the individual masses and luminosities to model LSPM J1314+1320AB, assuming solar composition and our standard CLES configuration (see Sect. \ref{std}). Given the A and B components are nearly equal in mass and luminosity, we took the averaged values  $\langle M \rangle=0.0881 \pm 0.0008 M_\odot$ and $\langle \log(L/L_\odot)\rangle = -2.623 \pm 0.010$. Results are presented in Table \ref{table:1-CLES_BHAC15}. The quoted errors are from propagating errors on measured mass and luminosity. More precisely, we computed various evolutionary tracks by varying observational constraints ($L$, $M$) within their given 1-$\sigma$ range and computed the respective 1-$\sigma$ confidence interval for output quantities $T_{\rm eff}$, age, $R$,  and log $g$. Table \ref{table:1-CLES_BHAC15} also shows results obtained with BHAC15 models, as determined by D16, which are similar. Note that, the derived $T_{\rm eff}$ is about 180 K hotter than that given by the spectral type-$T_{\rm eff}$ relation of \citet{Herczeg2014} used by D16, which gives $T_{\rm eff}=$ 2770 $\pm$ 100 K for  LSPM J1314 A and B.

\begin{deluxetable*}{ccCrlc}[b!]
\tablecaption{Stellar evolutionary models for LSPM J1314+1320AB}
\tablecolumns{3}
\tablenum{1}
\tablewidth{0pt}
\tablehead{
\colhead{} &
\colhead{CLES} &
\colhead{BHAC15}
}
\startdata
$T_{\rm eff}$ (K)  & $2950 \pm 6$ & $2950 \pm 4$  \\
age (Myr)    & $81.7  \pm 3.6$ & $80.8  \pm 2.5$  \\
log $g$         	& $4.840  \pm  0.013$ & $4.839  \pm  0.009$ \\
$R/R_\odot$   &  $0.1868 \pm  0.0021$ & $0.1871 \pm  0.0016$
\enddata
\tablecomments{\label{table:1-CLES_BHAC15} Results from evolutionary models for LSPM J1314+1320AB with reference mean values: $\langle M\rangle=0.0881 \pm 0.0008 M_\odot$ and $\langle \log(L_{bol}/L_\odot)\rangle = -2.623 \pm 0.010 $. Comparison between CLES and BHAC15 models, assuming solar metallicity.}
\end{deluxetable*}

\begin{figure}[!ht]
\begin{center}
\includegraphics[scale=0.4,angle=0]{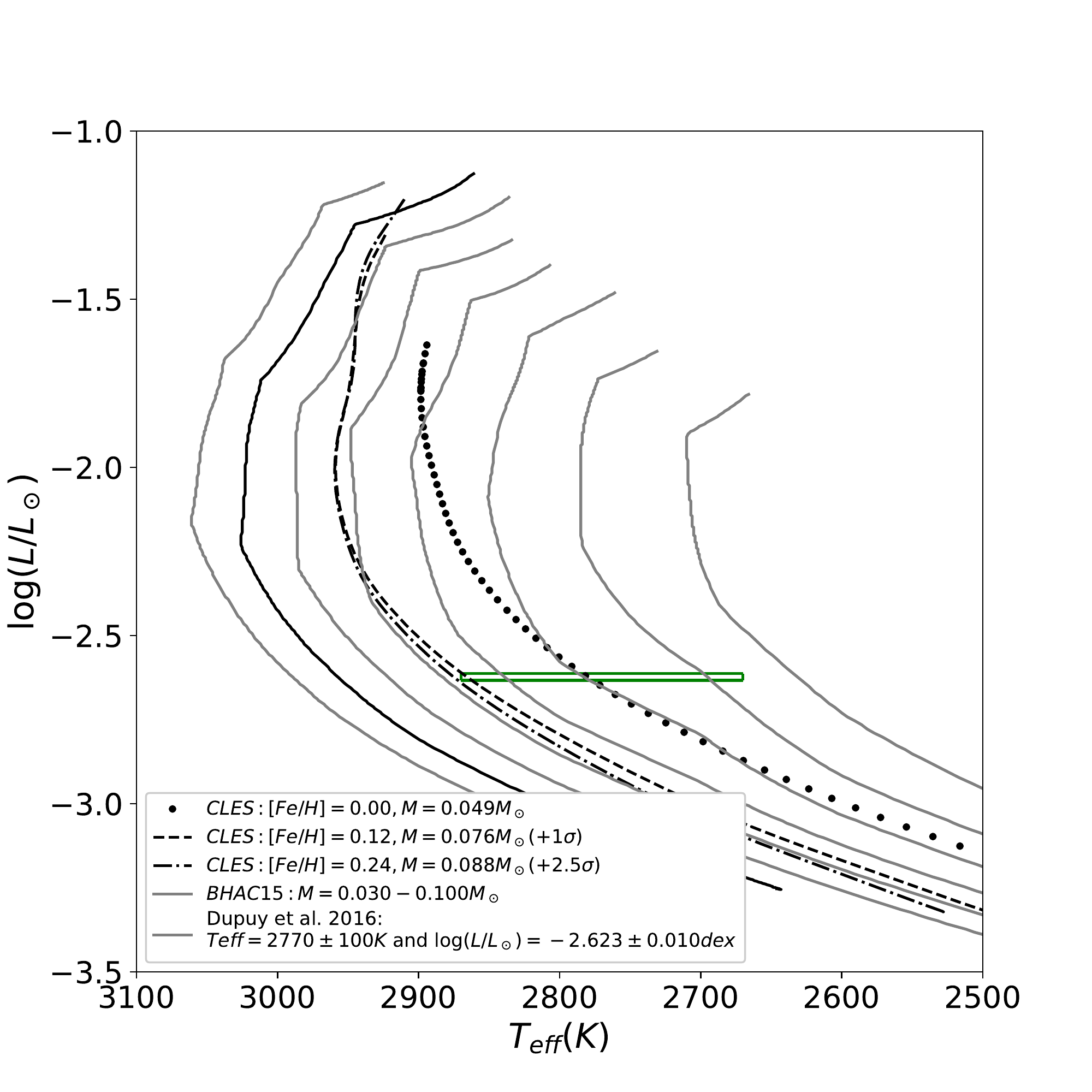}
\end{center}
\caption{\label{fig:3-HRdupuy2016} HR diagram comparison between UCD models and reference values for LSPM J1314+1320AB (Dupuy et al. 2016): $T_{\rm eff}=$ 2770 $\pm$ 100 K and $\log(L_{bol}/L_\odot)\rangle = -2.623 \pm 0.010$. BHAC15 tracks are from 0.03 to 0.10 $M_\odot$ by step of 0.01 $M_\odot$ and CLES tracks resulted from optimization process.} 
\end{figure}

In a second test, we reproduced the test proposed by D16 that mimicked the typical application of models by observers, i.e., that we have access to a measurement of the luminosity and to an effective temperature based on a spectral type$-T_{\rm eff}$ relation, but not to stellar masses. To carry out this test, we used our in-house Levenberg-Marquadt optimization algorithm (\citeauthor{press1992numerical} \citeyear{press1992numerical}, see \citeauthor{VanGrootel2018} \citeyear{VanGrootel2018} for details).
We first assumed solar composition and obtained a stellar mass $M=0.049 \pm 0.017 M_\odot$ (51$\pm18 M_{\rm Jup}$) with age $= 25 \pm 15$ Myr (Fig. \ref{fig:3-HRdupuy2016}, dashed black curve). Here again, quoted errors simply come from error propagation on $T_{\rm eff}$ and $L$. No systematic error (see Sect. \ref{syst}) was included. This result is consistent with the values from D16 interpolated from BHAC15 models: $M=50 ^{+20}_{-13} M_{\rm Jup}$ and age $= 25 ^{+10}_{-17}$ Myr, still showing the discrepancy between mass of stellar models and direct measurements. 
Adjusting the metallicity of LSPM J1314 to [Fe/H]=0.12 (+1$\sigma$) yields a higher mass $M=0.057 \pm 0.019 M_\odot$, closer to the measured value.
To fully reconcile models with observations, we found that we have to increase the metallicity to +2.5$\sigma$, i.e., [Fe/H]=0.24 (Fig. \ref{fig:3-HRdupuy2016}, dot-dashed green curve).
Increasing the metallicity to recover stellar parameters was also found in the study of TRAPPIST-1 \citep{VanGrootel2018}. In the same line, \citet{Lindgren2017} and \citet{Rajpurohit2018b} recently observed a typical average deviation of 0.2 to 0.4 dex in [Fe/H] between direct spectral fitting and calibration-based techniques.
One possibility is that metallicity is underestimated, another is residual error in models that can be 'adjusted' by artificially increasing metallicity. A third possibility is that the spectral type$-T_{\rm eff}$ relation depends on the scale in use: for a same spectral type, \cite{Herczeg2014}, \cite{Rajpurohit2013} and \cite{Passegger2018} show differences of 10\% in $T_{\rm eff}$ in the M-dwarf regime. 

\subsection{Proxima Centauri and GJ65 AB}
\label{sect3.2}
Proxima Centauri (GJ 551, M5.5Ve) is the nearest star after the Sun. It most likely part of a gravitationally bound triple system with $\alpha$ Centauri AB, with an orbital period of about 550 000 yr \citep{Kervella2017}. GJ65 AB (BL+UV Ceti, M5.5Ve+M6Ve) is a pair of red dwarfs in a sufficiently large binary to neglect gravitational and magnetic interactions, but sufficiently close to resolve the orbital motion within a human lifetime, with $P_{\rm orb}\sim 26.3$ yr (K16). These three nearby stars are among the most studied red dwarfs, and as such constitute cornerstones for models of very low mass stars. Although they are not strictly speaking UCDs, we can still confront our models to these unique benchmarks. 

\begin{figure}[h]
\begin{center}
\includegraphics[scale=0.4,angle=0]{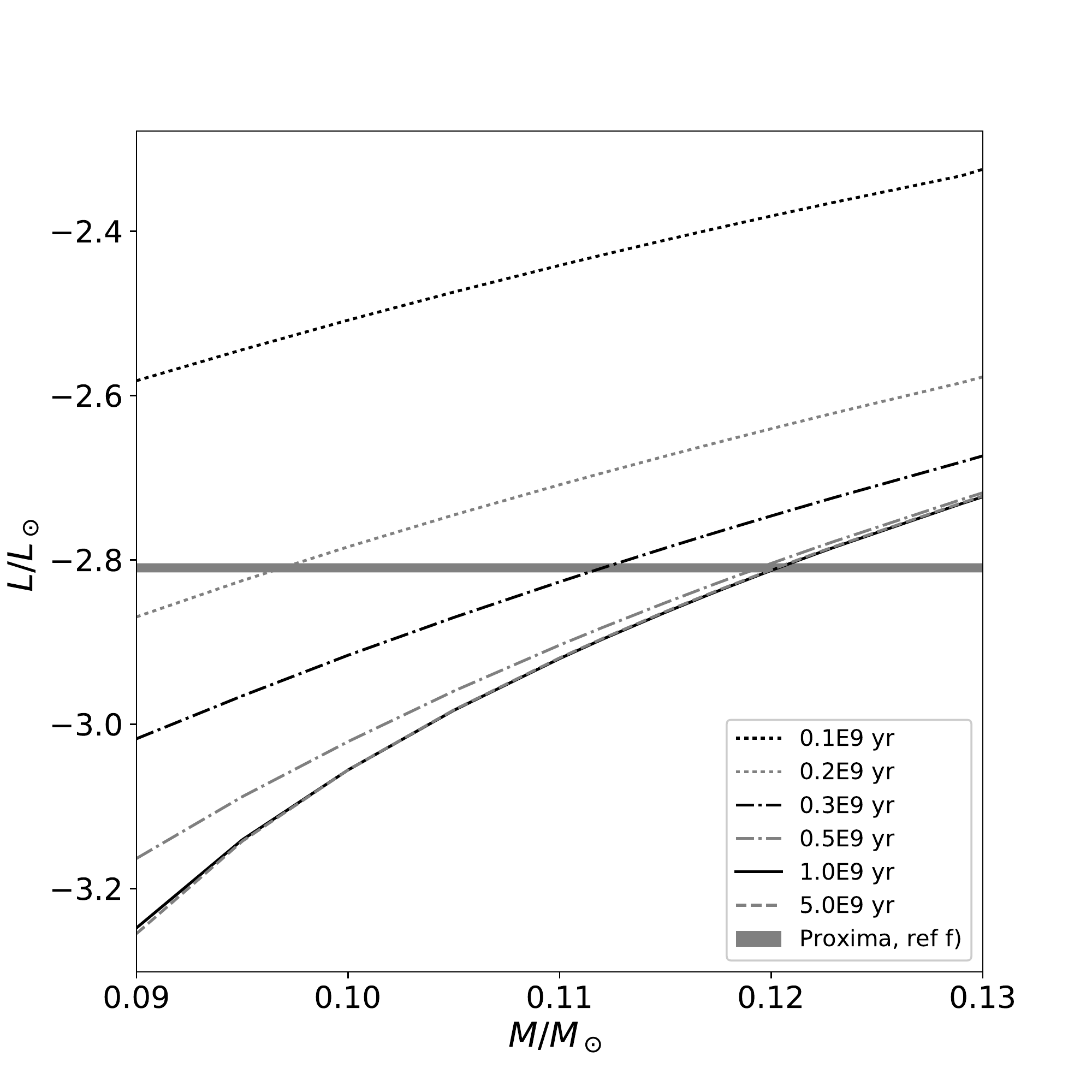}
  \includegraphics[scale=0.4,angle=0]{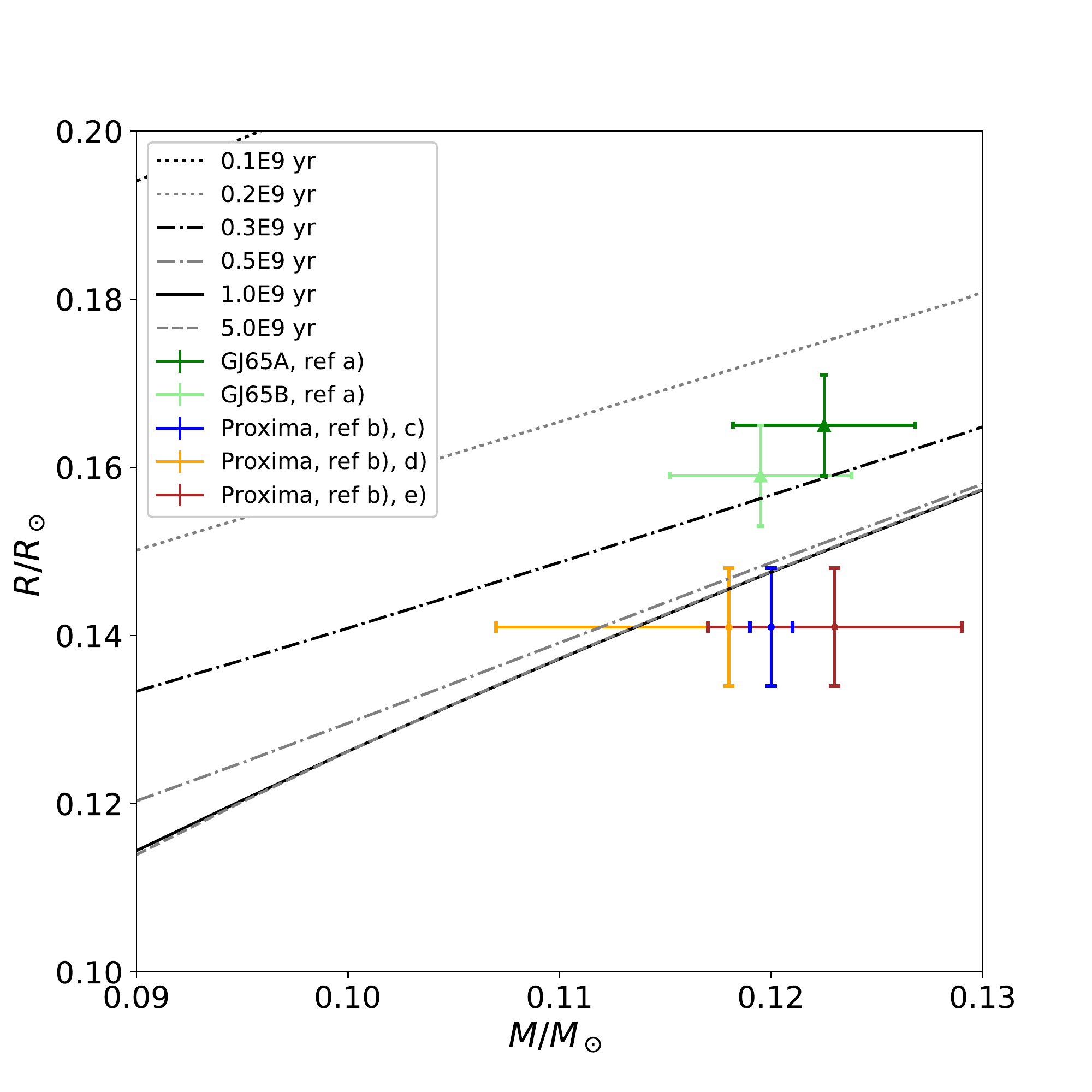}
    \end{center}
  \caption{\label{CLESMR} Mass-luminosity (left) and Mass-radius relation (right) for CLES models and for several isochrones (0.1-5.0 Gyr), assuming solar composition (see fig.10 in K16). Comparison to measurements for Proxima and GJ65 AB. References: a) \cite{Kervella2016}, b) \cite{Demory2009}, c) this work, d) \cite{Henry1999}, e) \cite{Delfosse2000}, f) \cite{Boyajian2012}.}
\end{figure}

Proxima and GJ65 AB all have three direct radii measurements from interferometry with, respectively, $0.141\pm0.007 R_\odot$ \citep{Demory2009}, $0.165\pm 0.006 R_\odot$ and $0.159\pm 0.006 R_\odot$ (K16). By orbital modeling, the total mass of GJ65 is accessible, as well as individual masses based on differential astrometry, indicating $0.1225\pm 0.0043 M_\odot$ and $0.1195\pm 0.0043 M_\odot$ for A and B components, respectively (K16). 
No direct mass measurement of Proxima exists, but it can be estimated through mass-absolute magnitude relations, $0.1239 \pm 0.0032 M_\odot$ using \citet{Mann2019} at solar-metallicity (distance and $K_S$ magnitude in \cite{Cutri2003}), $0.123\pm 0.006 M_\odot$ using \citet{Delfosse2000} relations, and $0.118 \pm 0.011 M_\odot$ using \citet{Henry1999}. 
The mass-luminosity relations obtained with CLES models are shown in fig. \ref{CLESMR} for solar-metallicity, giving us an estimated mass of $0.120 \pm 0.001 M_\odot$ for Proxima at 5 Gyr assuming a luminosity $L=0.00155 \pm 0.00002 L_\odot$ \citep{Boyajian2012}.
Directly determining metallicities from high-resolution spectroscopy of red dwarfs is a delicate task. For Proxima, iron abundance measurements range from [Fe/H]$=-0.07 \pm 0.14$ \citep{Passegger2016} to [Fe/H]$=+0.16 \pm 0.20$ \citep{Neves2014}. We could also consider that its membership in the $\alpha$ Cen system indicates a common origin from the same formation cloud, and thus the same age and initial composition. The metallicities of $\alpha$ Cen A and B have been accurately determined from atmospheric abundances of many elements \citep[e.g.][]{PortoMello2008}, and stellar modeling and asteroseismology allows us to obtain the initial composition and age, with $X_0\sim 0.70$, $Z_0 \sim 0.025$ and an age about 6 Gyr \citep{Bazot2016}. It is also possible that Proxima has been captured by $\alpha$ Cen, and thus does not share a common composition. Whatever the actual metallicity of Proxima, K16 adopted a differential approach to Proxima to determine the metallicity of GJ65 AB. Adopting [Fe/H]$=+0.05\pm0.20$ for Proxima, gave [Fe/H]$=-0.03\pm0.20$ and [Fe/H]$=-0.12\pm0.20$ for GJ65 A and B. 

K16 used the BHAC15 models (with solar composition) to model Proxima and GJ65 AB. They found that, while Proxima fits with the expected mass-radius relation, GJ65 AB both appear inflated, exceeding model expectations by 14$\pm$4\% and 12$\pm$4\%, respectively. This radius inflation could naturally be explained by a young age (about 250 Myr), but the GJ65 velocity vector likely indicates a star of at least 1 Gyr, and possibly much older. K16 carefully examined possible sources of discrepancy, and concluded that the most likely explanation is a reduced convection efficiency for GJ 65AB by a strong internal magnetic field linked to the relatively fast rotation of both stars ($v \sin i \sim$ 30 km s$^{-1}$). Proxima, on the contrary, is a slow rotator ($v \sin i \sim$ 2 km s$^{-1}$).

We modelled Proxima and GJ65 AB using the same optimization procedure as in the previous section. First, we placed GJ65 AB and Proxima in our CLES mass-radius (M-R) plot with 0.1-5.0 Gyr isochrones, for solar composition (Fig. \ref{CLESMR}).
The masses reference for Proxima plotted in Fig. \ref{CLESMR} are the values from \cite{Delfosse2000} and \cite{Henry1999} (red and orange, respectively) in which we also include our result (blue). We agree with the K16 conclusions: while Proxima reasonably fits the model expectations, GJ65 AB is either a young star (about 300 Myr), or somewhat inflated compared to its mass. 
In more detail, we found that models slightly \textit{overestimate} the stellar radius of Proxima, opposite to the usual trend (see, e.g., section \ref{sect3.1} and \citeauthor{VanGrootel2018} \citeyear{VanGrootel2018}). The situation would be worse considering Proxima to have the same metallicity as $\alpha$ Cen. 
We tested the suggestion of K16 to decrease the convection efficiency for GJ65 and found that we had to turn $\alpha_{\rm MLT}$ down to 0.03 (and even slightly lower for GJ65 A) in order to reconcile the measured radii with model expectations.
If we play with metallicity, we have to increase it by a great amount (more than [Fe/H]=+0.5) to reconcile interferometric and model radii. This is also unlikely given the differential spectroscopy carried out by K16, which indicates that Proxima and GJ65 have similar metallicities.

We reversed the problem by identifying which stellar mass would fit the radii measured from interferometry, assuming several Gyrs stars in both cases. For Proxima we found a mass of 0.113$\pm$0.007 $M_\odot$. This is within 2$\sigma$ of the value from \citet{Delfosse2000} relations, and within 1$\sigma$ of the \citet{Henry1999} ones. Proxima is at the limit of validity for applicability of such relations (11, 10, and 9.5 for the $JHK$ absolute magnitudes, respectively, for \citealt{Delfosse2000}). Indeed, applying such relations for GJ65 (also at the validity limit), K16 found that predicted masses are 12\% and 17\% lower than the measured masses, respectively. Repeating the same exercise with GJ65, we found that we are within 2$\sigma$ of the dynamical value: a mass of 0.137$\pm$0.006 $M_\odot$ is needed to account for the radius of 0.165$\pm$ 0.006 $R_\odot$ (GJ 65A), and a mass of 0.132$\pm$0.006 $M_\odot$ is needed to account for the radius of 0.159$\pm$ 0.006 $R_\odot$ (GJ 65B). 

In conclusion, given uncertainties in stellar models in metallicity and $\alpha_{MLT}$ for stars with inhibit convection (e.g., fast rotating or strong magnetic field, see review in \citeauthor{brun2017magnetism} \citeyear{brun2017magnetism}), and the uncertainties that are likely to occur in mass-absolute magnitude relations, we posit that an agreement within 2$\sigma$ is acceptable.


\section{Conclusion}
\label{sect4}

We have presented new evolutionary models for  $T_{\rm eff} \geq$ 2000 K UCD objects, which encompass the very low-mass stars and young, still contracting brown dwarfs. These models are based on our in-house evolutionary code CLES adapted to produce UCD models. In particular, we included a relevant EOS that includes H, He, C and O elements, and appropriate boundary conditions from two sets of model atmospheres. We presented some properties of our models and investigated their systematic error associated with uncertainties in input physics. We showed a comparison with the reference BHAC15 models in the UCD regime. Finally, a series of test-cases was carried out to visualize the strengths and limits of our models.

Tables for our CLES models for various UCD masses and metallicities can be found at \url{http://www.astro.ulg.ac.be/ASTA/cles-models-UCDs/} 

\begin{acknowledgements}
The authors thank the anonymous referee for an helpful review. We warmly thank Didier Saumon for various advices when developing our UCD models. We thank the whole SPECULOOS team (\url{http://www.speculoos.uliege.be/}) for motivation, advice and encouragement when developing these models. We thank Uffe G. J{\o}rgensen for his comments after careful reading of the manuscript. C.S.F. is funded by an Action de Recherche Concert\'ee (ARC) grant financed by the Wallonia-Brussels Federation. V.V.G. is F.R.S.-FNRS Research Associate. The research leading to these results has received funding from the ARC grant for Concerted Research Actions, financed by the Wallonia-Brussels Federation. B.A. was supported by the ERC Consolidator Grant funding scheme ({\em project STARKEY}, G.A. n.~615604). G.F. acknowledges the contribution of the Canada Research Chair Program.
\end{acknowledgements}

\software{CLES \citep{Scuflaire2008}}

\bibliography{mybib}



\end{document}